\documentclass[aps, prb, amsmath,amssymb,
 reprint, 
 superscriptaddress,
]{revtex4-2}

\usepackage{graphicx}
\usepackage{dcolumn}
\usepackage{bm}
\usepackage{color}
\usepackage{longtable}
\usepackage[colorlinks=true,allcolors=blue]{hyperref}
\usepackage{ulem}

\begin{document}

\preprint{APS/123-QED}

\title{Thermal Induced Structural Competitiveness and Metastability of Body-centered Cubic Iron under Non-Equilibrium Conditions}

\author{Shuai Zhang}
\email{szha@lle.rochester.edu}
\affiliation{Laboratory for Laser Energetics, University of Rochester, Rochester, New York 14623, USA}

\author{Aliza Panjwani}
\affiliation{Laboratory for Laser Energetics, University of Rochester, Rochester, New York 14623, USA}
\affiliation{Department of Chemistry, University of Rochester, Rochester, New York 14611, USA}

\author{Penghao Xiao}
\affiliation{Department of Physics and Atmospheric Science, Dalhousie University, Halifax, Nova Scotia B3H 4R2, Canada}

\author{Maitrayee Ghosh}
\affiliation{Laboratory for Laser Energetics, University of Rochester, Rochester, New York 14623, USA}
\affiliation{Department of Chemistry, University of Rochester, Rochester, New York 14611, USA}

\author{Tadashi Ogitsu}
\affiliation{Lawrence Livermore National Laboratory, Livermore, California 94550, USA}

\author{Yuan Ping}
\affiliation{Lawrence Livermore National Laboratory, Livermore, California 94550, USA}

\author{S. X. Hu}
\affiliation{Laboratory for Laser Energetics, University of Rochester, Rochester, New York 14623, USA}
\affiliation{Department of Physics and Astronomy, University of Rochester, Rochester, New York 14627, USA}
\affiliation{Department of Mechanical Engineering, University of Rochester, Rochester, New York 14627, USA}

\date{\today}

\begin{abstract}
The structure and stability of iron near melting at multi-megabar pressures are of significant interest in high pressure physics and earth and planetary sciences.
While the body-centered cubic (BCC) phase is generally recognized as unstable at lower temperatures, its stability relative to the hexagonal close-packed (HCP) phase at high temperatures (approximately 0.5~eV) in the Earth's inner core (IC) remains a topic of ongoing theoretical and experimental debate.
Our {\it ab initio} calculations show a significant drop in energy, the emergence of a plateau and a local minimum in the potential energy surface, and stabilization of all phonon modes
at elevated electron temperatures ($>$1--1.5 eV).
These effects increase the competition among the BCC, HCP, and the face-centered cubic (FCC) phases and lead to the metastability of the BCC structure.
Furthermore, the thermodynamic stability of BCC iron is enhanced by its substantial lattice vibration entropy.
This thermally induced structural competitiveness and metastability 
under non-equilibrium conditions provide a clear theoretical framework for understanding iron phase relations and solidification processes, both experimentally and in the IC.
\end{abstract}

\maketitle


{\it Introduction.} The Earth possesses a young solid center known as the inner core (IC), which formed within a liquid dominated by iron. Solidification occurred when the melting curve of the iron alloy intersected with the adiabatic temperature-depth profile of the liquid during the cooling of the core. The nucleation and growth of the IC are believed to have supplied a crucial energy source that revitalized the geodynamo, which otherwise would have ceased $\sim$0.5 billion years ago~\cite{cormier_book_chap8_2021, bono_young_2019, driscoll_geodynamo_2019, zhou_early_2022}.

The nucleation (liquid to solid) or melting (solid to liquid) of iron is therefore essential for understanding the geodynamo and many seismological observations of the Earth's deepest interior. However, this has been a perplexing topic. Numerous experiments and theories have indicated that the hexagonal close-packed (HCP) phase is the most stable below the melting temperature ($T_\text{m}$)~\cite{ping_solid_2013, kraus_measuring_2022, sun_two-step_2022, gonzalez-cataldo_ab_2023, li_competing_2024}. Yet, there have also been reports of stability or metastability of the body-centered cubic (BCC) phase in iron and its alloys near $T_\text{m}$ at high pressures~\cite{DubrovinskyScience2007, belonoshko_stabilization_2017, sun_two-step_2022, ChatterjeeMinerals2021, ghosh_cooperative_2023, li_deep-learning_2024}. 
Moreover, BCC as a candidate to the structure of the IC is attractive because this would offer explanations to several seismological observations, such as positive correlations between P wave velocity and attenuation and a high P-wave anisotropy required to explain the correlation~\cite{ghosh_cooperative_2023}.

Despite vast advances in high-pressure experimental and computational technologies in recent years, the multi-megabar pressures and thousands-Kelvin temperatures still pose extreme challenges to nailing down the structure and properties of iron under the conditions of Earth's IC. 
Our recent {\it ab initio} calculations~\cite{ghosh_cooperative_2023} show that, in addition to its increased dynamical stability (when the electron-ion equilibrium temperature approaches $\sim$1000--2000~K below the melting point), BCC iron can also become mechanically stable through electron thermal effects alone (when the electron temperature $T_\text{e}$ exceeds 12,000 K, or approximately 1~eV). 
These suggest iron's potential energy surface (PES) changes and facilitates stabilization of BCC iron at high temperatures.

In this Letter, we perform further calculations from first principles and provide new insights into the PES change, transition pathway, and free energy of iron.
Our findings aim to establishing a solid foundation for the growing body of research in the community addressing the perplexing problem of iron’s structure and BCC phase stability under conditions relevant to both the Earth’s inner core (IC) and cutting-edge laser and x-ray-driven high-pressure facilities.

{\it Methodology.} We perform {\it ab initio} calculations based on Mermin--Kohn--Sham finite-temperature density functional theory (DFT), as implemented in Vienna {\it ab initio} simulation package (VASP)~\cite{kresse96b}. 
The electron temperature $T_\text{e}$ is set by enforcing Fermi--Dirac distribution of the electrons in their eigenstates according to their energies.
These calculations include PES, transition pathway, and phonon at $T_\text{e}$ between 6,000 and 30,000 K.
The lower and the upper bounds of $T_\text{e}$ are relevant to temperatures of the IC and in the regime of mechanical stability of BCC iron as shown in Fig.6a of~\cite{ghosh_cooperative_2023}, respectively. 
Three structures have been considered: BCC, hexagonal close-packed (HCP), and face-centered cubic (FCC).

We calculate the PES under the constraint of four-atom orthorhombic cells with static ions and a fixed volume of 7.2 \AA$^3$/atom, 
same as that considered in~\cite{godwal_stability_2015}.
For calculating the transition pathways, we employ the solid-state nudged elastic band (SS-NEB) method~\cite{sheppard_generalized_2012, ghasemi_nudged_2019},
which treats the cell degrees of freedom and atomic positions on an equal footing.
In the SS-NEB calculations, we use four-atom unit cells~\footnote{Atoms fractional coordinates of the four-atom cells used in the calculations in this study are
$(0.0, 0.25, 0.25)$, $(0.0, -0.25, 0.75)$, $(0.5, 0.75, 0.25)$, $(0.5, 0.25, 0.75)$ for BCC and FCC, 
$(0.0, 0.33, 0.25)$, $(0.0, -0.33, 0.75)$, $(0.5, 0.83, 0.25)$, $(0.5, 0.17, 0.75)$ for HCP,
and geometries of the cells are
$(\sqrt{2}, \sqrt{2})$, $(1, 1)$, $(\sqrt{3}, \sqrt{8/3})$ for the ratios of lattice constants $(b/a, c/a)$ in BCC, FCC, and HCP, respectively.}
for the end-member structures (BCC, FCC, and HCP), dense ($10\times10\times10$ or finer, depending on the cell shape) $\Gamma$-centered $k$ mesh to sample the Brillouin zone, and find the pathways by relaxing all structures along the reaction coordinates until internal energies and atomic forces converge to better than 1 meV and 0.01 eV/\AA, respectively.
We use a well-established technique to carry out phonon calculations under the quasi-harmonic approximation (QHA):
first, we compute the force constants using density-functional-perturbation theory (DFPT) as implemented in VASP;
second, we use PHONOPY~\cite{phonopy} to calculate the phonon frequencies along standard high-symmetry $q$ paths of the corresponding cell~\cite{seek-k_path} and thermodynamic properties by using a $10\times10\times10$ $q$ mesh.
In the DFPT stage, we use 16 or 32 atom cells and $\Gamma$-centered $6\times6\times6$, or finer $k$ mesh.
Both SS-NEB and phonon calculations are performed at four different $T_\text{e}$ conditions (6,000, 12,000, 18,000, and 30,000 K)
and two different pressures (220 GPa and 317.6 GPa, 
which are approximately 230 GPa, reachable in diamond-anvil-cell experiments, and 330 GPa, same as that at Earth's IC boundary,
when adding the ion thermal contributions at 6,000~K, respectively).

All calculations use the projector augmented-wave type~\cite{Blochl1994} Fe\_pv potential (with core radius of 2.20 Bohr and 14 outermost electrons treated as valence), Perdew--Burke--Ernzerhof exchange-correlation functional~\cite{Perdew96}, and 400 eV cutoff for the plane-wave basis set.
These settings have been carefully tested in~\cite{ghosh_cooperative_2023} and chosen for calculations of iron at IC conditions. 
Our tests show the magnetic moment $M$ of iron becomes negligible at the IC pressures.
In addition, because $M$ decreases with temperature~\cite{han_phonon_2018}, 
electron spin can be ignored at the high-pressure, temperature conditions considered in this study.

\begin{figure}[ht]
\centering
\includegraphics[width=1.0\linewidth]{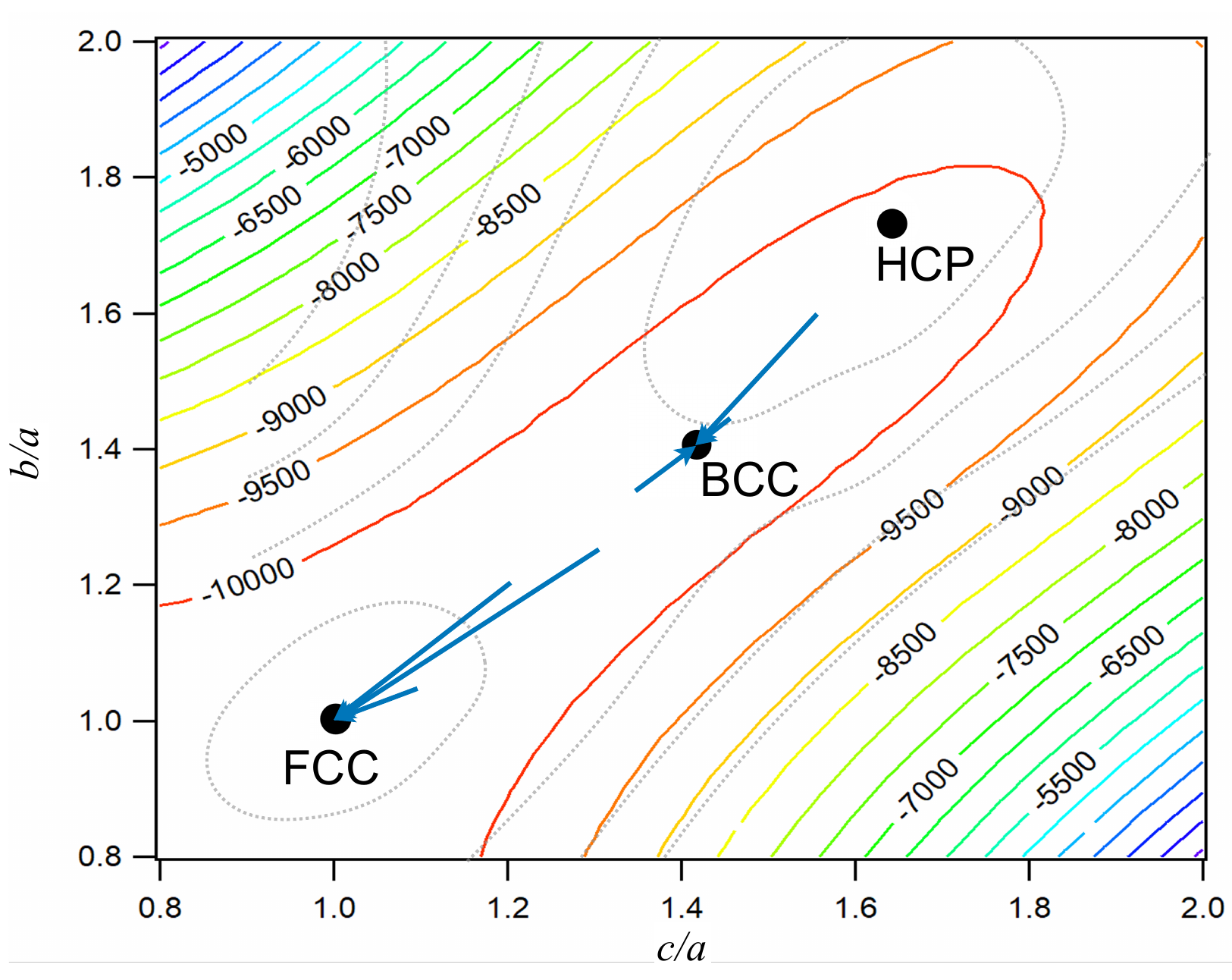}
\caption{Contours (in 500-meV interval) of the potential energy surface of iron in a 2D orthorhombic distortion space, based on isochoric four-atom unit cells, at electronic temperature $T_\text{e}$=18000 K and volume of 7.2\AA$^3$/atom. Previously calculated PES contours at 0 K and the same density by Godwal {\it et al.}~\cite{godwal_stability_2015} is shown in gray dotted curves for comparison. Blue arrows indicate structural relaxation calculations at $T_\text{e}$=18,000 K with the tail and head of each arrow representing the initial and final structures, respectively. 
Energies labeled are in units of meV/atom.
}
\label{fig1pes}
\end{figure}

{\it Results.}
First, we demonstrate the changes in the PES of iron with temperature.
For this purpose, we consider a cell-distortion space that is characterized by two variables $c/a$ and $b/a$, which are the ratios of cell parameters, while keeping the cell volume and atomistic coordinates to be fixed.
This approach has been employed by Godwal {\it et al.}~\cite{godwal_stability_2015} to study iron at the same density but 0 K,
which shows a double-basin feature [around FCC and HCP~\footnote{We note that, strictly speaking, the HCP structure does not fall into the same 2D cell-distortion space for FCC and BCC, because the atomistic coordinates are different, but we choose to follow that in Ref.~\cite{godwal_stability_2015} and still labeled HCP in Fig.~\ref{fig1pes} by using the ratios of its cell parameter.} (see gray dotted ovals in Fig.~\ref{fig1pes})].
Our new results at $T_\text{e}=18,000$~K show a dramatic difference---the two basins merge into one large envelope (the red -10 eV contour line) that includes BCC. 
Within this regime, structures have smaller differences in energy than at 0 K.
This finding is corroborated by our additional relaxation calculations, 
which show a number of perturbed structures near the BCC structure relax back into BCC [and similarly for FCC (see blue arrows in Fig.~\ref{fig1pes})].
This suggests a potential well emerges around the BCC structure at such high temperatures, which does not exist at 0 K.
We note this is consistent with the conditions of mechanical stability (defined by the relation in elastic constants $C_\text{11}>C_\text{12}$) of BCC iron at $T_\text{e}>12,000$~K as found in~\cite{ghosh_cooperative_2023}.

\begin{figure}[ht]
\includegraphics[width=1.0\linewidth]{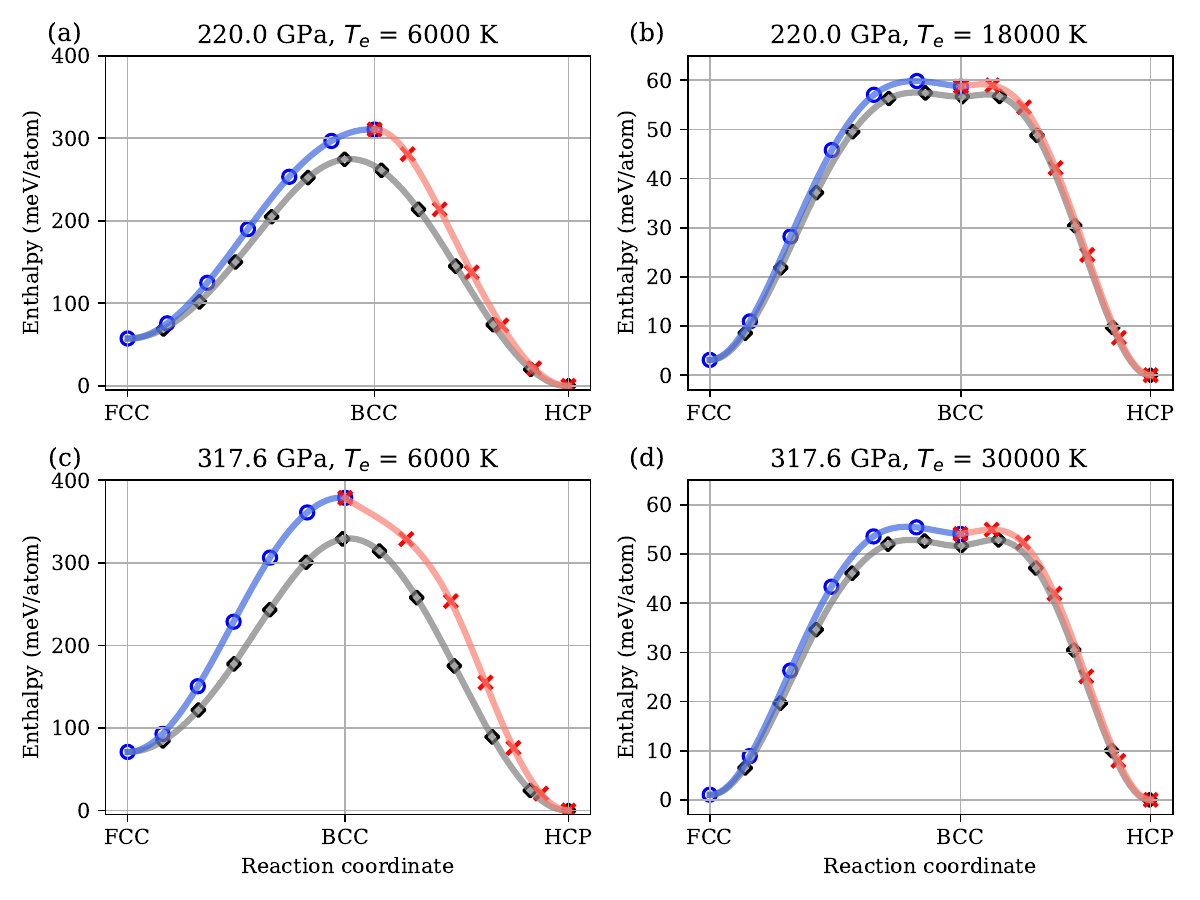}
\caption{Enthalpy along NEB transition pathways of HCP, FCC, and BCC iron at different pressure and $T_\text{e}$ conditions. 
In each plot, the enthalpy is shown relative to that of the HCP structure.}
\label{fig2barrier}
\end{figure}

Next, we calculate the changes in the PES with $T_\text{e}$ more systematically using the SS-NEB approach.
In the SS-NEB calculations, both the cell shape and atomistic positions change along the pathways FCC--HCP and BCC--HCP, while only the cell shape varies for FCC--BCC (see supplementary materials~\cite{urlsuppmat}).
Figure~\ref{fig2barrier} compares enthalpy (relative to HCP) as a function of reaction coordinates along the transition paths at two different pressures
and electron temperatures $T_\text{e}$.
At relatively low $T_\text{e}$ (6,000 K), HCP has the lowest enthalpy, followed by FCC (higher than HCP by approximately 70 meV/atom),
while the enthalpy of BCC is higher than other structures along the NEB paths, including the saddle point along the FCC--HCP path.
This indicates BCC is unstable and a barrier $\Delta H$ of approximately 300--400 meV/atom, larger at higher pressure, needs to be overcome for the FCC$\leftrightarrow$HCP transition.
With increasing $T_\text{e}$, the differences between all pairs of structures become smaller (e.g., FCC higher than HCP by 5 meV/atom or less), $\Delta H$ decreases, and the enthalpy profile flattens around BCC.
At $T_\text{e}$ of 18,000--30,000~K, $\Delta H$ for FCC$\leftrightarrow$HCP is approximately 50--60 meV/atom, and a sub-meV/atom enthalpy barrier  develops for BCC transitioning to the close-packed structures. 

A closer investigation shows that the NEB path for FCC$\leftrightarrow$HCP (gray curve in Fig.~\ref{fig2barrier}) transitions through BCC, and the small differences from the FCC--BCC (blue) and BCC--HCP (red) paths are due to removal of symmetry constraint in the FCC--HCP calculations.
This results in structures at the minimum of the plateau region to have $c/a$ and $b/a$ that are slightly different from the ideal value ($\sqrt{2}$) and principal stress components slightly anisotropic (both by less than {\color{black}$0.5\%$}), but space group remains the same ($Im\bar{3}m$) as that of ideal BCC.
These observations confirm the existence of a basin structure in the PES around BCC at high temperature, as suggested by the energy contour changes and relaxation findings (Fig.~\ref{fig1pes}).
Moreover, these reflect the faster rate of decreasing in enthalpy of BCC with $T_\text{e}$ than other structures, enhancing its relative stability at high $T_\text{e}$.

Finally, we consider the effect of only atomistic displacements by performing phonon calculations for HCP, FCC, and BCC iron at the same pressures and $T_\text{e}$ as those considered in the SS-NEB calculations.
At these conditions, we find all phonon modes of HCP and FCC are stable.
In contrast, BCC shows imaginary frequencies along the $H-P$ (i.e., $\{\xi\xi\xi\}$ in displacement) direction at $T_\text{e}=6,000$~K, but all frequencies become real at $T_\text{e}\ge12000$~K, suggesting BCC becomes dynamically stable at the high $T_\text{e}$ conditions.

\begin{figure}[ht]
\centering
\includegraphics[width=1.0\linewidth]{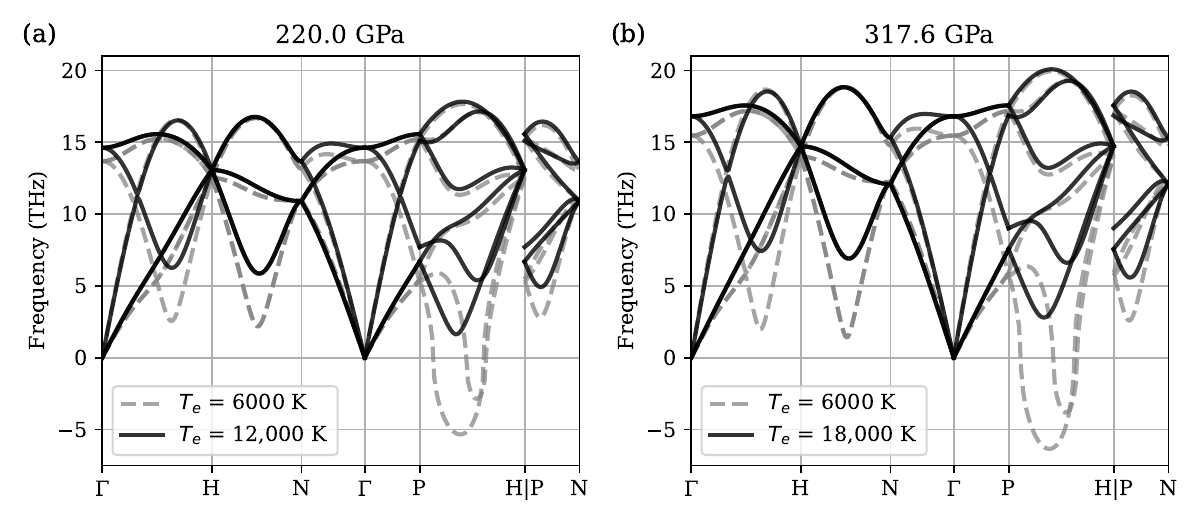}
\caption{Phonon band structures of BCC iron at different electron temperatures $T_\text{e}$ and (a) 220 GPa and (b) 317.6 GPa, excluding ion thermal contributions (if these contributions are included, the pressures would increase to approximately 230 and 330 GPa, respectively, at 6,000 K). 
Negative values at $T_\text{e}$=6,000 K denote imaginary frequencies.}
\label{fig3phonon}
\end{figure}

\begin{figure}[ht]
\centering
\includegraphics[width=1.0\linewidth]{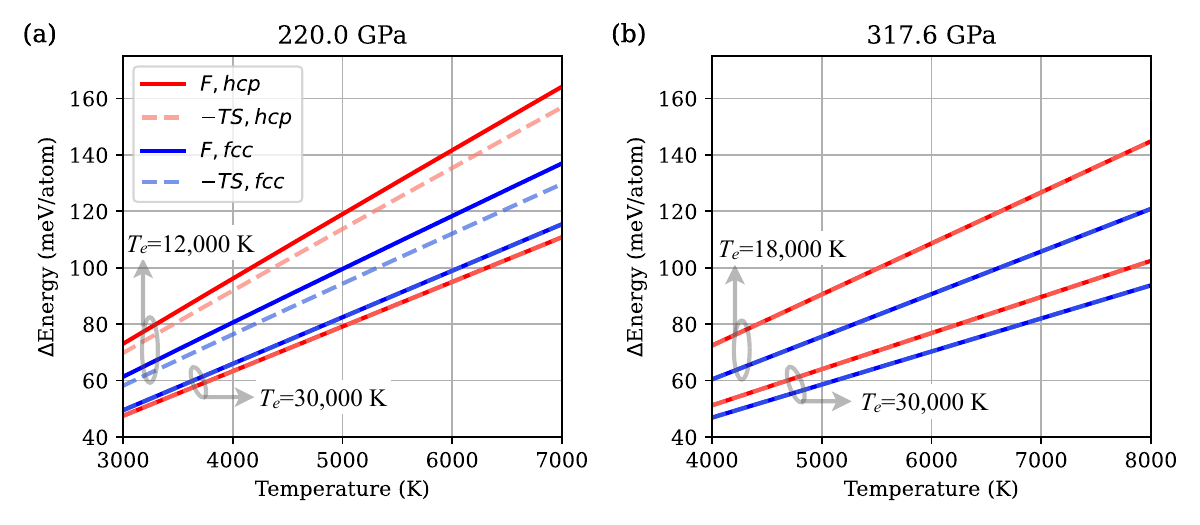}
\caption{Vibrational free energy ($F$) differences of HCP and FCC compared to BCC iron at different ion temperatures based on the quasi-harmonic phonon results at (a) 220 GPa and (b) 317.6 GPa, excluding ion thermal contributions (if these contributions are included, the pressures would increase to approximately 230 and 330 GPa, respectively, at 6,000 K).
Corresponding differences in vibrational entropy contributions ($-TS$) are also shown for comparison, which are slightly lower than $F$ at $T_\text{e}$=12,000 K in (a) and overlap with the $F$ curves in all other conditions.}
\label{fig4Fvib}
\end{figure}

The phonon results allows unambiguous estimation of the lattice vibration (or ``ion thermal'', using typical equation-of-state language) contribution to the thermodynamic equation of state 
(the electron thermal contributions of the three phases at each $T_\text{e}$ are expected to be the same because of their similarity in electronic structure~\cite{pourovskii_electronic_2020}).
We have thus calculated the vibrational entropy and free energy
and compared the results (relative to BCC) in Fig.~\ref{fig4Fvib} at ion temperatures $T_\text{i}$ of approximately $\pm2,000$ around the melting point ($\sim5,000-6,000$~K at 230 GPa or $\sim6,000-7,000$~K at 330 GPa).
The results indicate that the vibrational free energy of BCC is significantly lower than that of FCC and HCP, owing to its larger vibrational entropy.
This relationship contrasts with the enthalpy trends (illustrated in Fig.~\ref{fig2barrier}).
The vibration-induced reduction in the free energy of BCC relative to HCP and FCC is 50--160 meV/atom, 
with greater effects observed at lower $T_\text{e}$ or higher $T_\text{i}$.
This reduction is comparable to the enthalpy differences of 50--400 meV/atom, 
effectively reconciling these values and resulting in the total free energy of BCC becoming lower than that of HCP and FCC, particularly at sufficiently high $T_\text{i}$.

It is also worthy to note that the vibrational free energy of FCC is lower than HCP (except for $T_\text{e}$=30,000 K at 230 GPa) due to their difference in entropy. This could make the total free energies of FCC and HCP to also become similar and more competitive to each other.


{\it Conclusions and discussions.}
Our results highlight the importance of non-equilibrium conditions with thermal electrons in affecting the dynamic and thermodynamic stability of iron.
In particular, with $T_\text{e}>$12,000--18,000~K (or 1--1.5~eV), the PES around the BCC phase shifts from a saddle point to a shallow well and all imaginary-frequency phonon modes disappear, which allows iron to become dynamically stable.
Furthermore, the larger vibrational entropy of BCC can reduce its free energy and make it thermodynamically competitive to HCP and FCC if temperature is sufficiently high (e.g., under superheated conditions).

Implications of the above results to the Earth's IC are twofold. First, a possible IC temperature of 6,000~K corresponds to a thermal kinetic energy of approximately 0.5 eV, which is larger than the enthalpy barrier between FCC and HCP ($\sim$0.3 eV, as shown in Fig.~\ref{fig2barrier}c).
The associated reaction rate, estimated by using the Eyring equation, is more than 400 times/picosecond. 
This implies iron can have substantial coexistence of FCC and HCP.
Second, the high temperature conditions can stabilize the thermal phonon modes of BCC iron (as previously shown in DFT-based molecular dynamics simulations~\cite{ghosh_cooperative_2023}),
whose large entropy relative to FCC and HCP can enhance the possibility for it to be metastable thermodynamically.
However, because of the similarity in total free energy of the different structures, anharmonic effects beyond what has been included within QHA still have to be rigorously quantified (to an accuracy of 1 meV) in order to generate a more complete picture about the stability of BCC iron.
This nevertheless is extremely challenging at such high temperatures as near melt, even for magnesium oxide which has a relatively simple phase diagram and electronic structure~\cite{zhang_toward_2023}.

Recently, important progress toward calculating free energies in meV accuracy has been made based on fully {\it ab initio} thermodynamic integration (TDI)~\cite{gonzalez-cataldo_ab_2023}.
However, due to the high complexity of methodology, intensity of computation, and lack of flexibility in {\it ab initio} codes, such high-accuracy TDI results for BCC and other phases at the same thermodynamic conditions are still missing. 
Large-scale classical molecular dynamics (CMD) codes typically offer more flexibility and easy computation in various ensembles. 
CMD with machine-learning or semi-empirical potentials with {\it ab initio} quality have shown promises toward reaching the ultimate goal of accurately simulating iron and calculating its free energies~\cite{sun_ab_2023, sun_unveiling_2024, li_deep-learning_2024, li_competing_2024}.
However, considering that the error associated with carefully trained potentials can still exceed 10$^2$ meV~\cite{wilson_can_2023}, extreme care and attention need to be paid to training the potentials and demonstrating their suitability to study such phase transitions with meV barriers as those shown in this study. 

Connecting to experiments,
the highly non-equilibrium condition between electrons and ions can be created in laboratory by ultrafast heating using optical laser pulses~\cite{fernandez-panella_reduction_2020} or x-ray pulses~\cite{mckelvey_thesis_2019}. 
The equilibration time is typically a few picoseconds or shorter at solid density and above. 
Although past experiments on compressed iron at nanosecond time scale did not show evidence of the BCC phase~\cite{ping_solid_2013}, combining the ultrafast heating with dynamic or static compression can reach the pressure and non-equilibrium conditions explored in this Letter and test our predictions.

\begin{acknowledgments}
This material is based upon work supported by the Department of Energy
[National Nuclear Security Administration] University of Rochester “National
Inertial Confinement Program” under Award Number DE-NA0004144.

This report was prepared as an account of work sponsored by an agency of the
US Government. Neither the US Government nor any agency thereof, nor any
of their employees, makes any warranty, express or implied, or assumes any
legal liability or responsibility for the accuracy, completeness, or
usefulness of any information, apparatus, product, or process disclosed, or
represents that its use would not infringe privately owned rights. Reference
herein to any specific commercial product, process, or service by trade name,
trademark, manufacturer, or otherwise does not necessarily constitute or imply its endorsement, recommendation, or favoring by the US Government or any
agency thereof. The views and opinions of authors expressed herein do not
necessarily state or reflect those of the US Government or any agency
thereof.

\end{acknowledgments}





%

%

\end{document}